\documentclass[12pt]{article} 

\usepackage{latexsym} 
\usepackage{amssymb}  
\usepackage{graphicx}       
\usepackage{lscape}

\newcommand{\Si}{\Sigma}

\newcommand\eqn[1]{(\ref{#1})}      

\newcommand{\beq}{\begin{equation}}
\newcommand{\eeq}{\end{equation}}
\newcommand{\ba}{\begin{array}}
\newcommand{\bea}{\begin{eqnarray}}
\newcommand{\ea}{\end{array}}
\newcommand{\eea}{\end{eqnarray}}

\newcommand\comment[1]{ \hbox{[{\it Comment suppressed here.}\/]} }
\newcommand\hide[1]{}

\newcommand{\Tr}{\hbox{Tr}}

\newcommand{\bp}{{\bf p}}
\newcommand{\bq}{{\bf q}}
\newcommand{\bk}{{\bf k}}

\newcommand{\skipover}[1]{}

\newcommand{\pa}{\parallel}
\newcommand{\pe}{\perp}

\makeatletter 

\def\appendix{\par                              
    \setcounter{section}{0}                     
    \setcounter{subsection}{0}
    \renewcommand{\theequation}{\Alph{section}.\arabic{equation}}
    \renewcommand{\thesection}{Appendix \Alph{section}
                \setcounter{equation}{0}  } 
}

\def\applabel#1{\@bsphack
  \protected@write\@auxout{}%
         {\string\newlabel{#1}{{\Alph{section}}{\thepage}}}%
  \@esphack}

\def\section{
\setcounter{equation}{0}        
\@startsection {section}{1}{\z@}{-3.5ex plus -1ex minus 
 -.2ex}{2.3ex plus .2ex}{\large\bf}}
\renewcommand{\theequation}{\arabic{section}.\arabic{equation}}

\def\subsection{\@startsection{subsection}{2}{\z@}{-3.25ex plus -1ex minus 
 -.2ex}{1.5ex plus .2ex}{\normalsize\bf}}

\def\subsubsection{\@startsection{subsubsection}{3}{\z@}{-3.25ex plus
 -1ex minus -.2ex}{1.5ex plus .2ex}{\normalsize}}

\makeatother   

\newsavebox{\eqlabel}

\makeatletter  
\newlength{\numblen}
\newsavebox{\eqnumb}
\def\@eqnnum{\savebox{\eqnumb}{\rm (\theequation)}%
\settowidth{\numblen}{\usebox{\eqnumb}}%
\makebox[\numblen][l]{\usebox{\eqnumb}~~~\usebox{\eqlabel}}}
\makeatother   

\newenvironment{equationwithlabel}[1]{ %
  \begin{equation}\label{#1} }{\end{equation}} 
\newcommand{\beql}[1]{\begin{equationwithlabel}{#1}}
\newcommand{\eeql}{\end{equationwithlabel}}

\begin{document}

\title{Critical phenomena from the two-particle irreducible 
$1/N$ expansion}

\author{
Mark Alford\thanks{email: alford@wustl.edu} $^a$,
\addtocounter{footnote}{1}
J\"urgen Berges\thanks{email: j.berges@thphys.uni-heidelberg.de} $^b$
and
Jack M.~Cheyne\thanks{email: j.cheyne@physics.gla.ac.uk} $^c$
\\
[2.ex]
\normalsize{$^a$ Physics Department, Washington University CB 1105}\\[0.ex]
\normalsize{One Brookings Drive, St.~Louis, MO 63130, USA}\\[1.ex]
\normalsize{$^b$ Institut f{\"u}r 
Theoretische Physik, Universit{\"a}t Heidelberg}\\[0.ex]
\normalsize{Philosophenweg 16, 69120 Heidelberg, Germany}\\[1.ex]
\normalsize{$^c$ Department of Physics and Astronomy, University
of Glasgow}\\[0.ex]
\normalsize{University Avenue, Glasgow, G12 8QQ, UK}
}


\date{}

\begin{titlepage}
\maketitle
\def\thepage{}          

\begin{abstract}
The $1/N$ expansion of the two-particle irreducible (2PI)
effective action is employed to compute universal 
properties at the second-order phase transition of an
$O(N)$-symmetric $N$-vector model directly in three dimensions.
At next-to-leading order the approach cures the spurious 
small-$N$ divergence of the standard (1PI) $1/N$ expansion
for a computation of the critical anomalous dimension $\eta(N)$,
and leads to improved estimates already for moderate values
of $N$. 
\end{abstract}

\end{titlepage}

\renewcommand{\thepage}{\arabic{page}}


\section{Introduction}

New methods for the quantitative description of thermal equilibrium
as well as nonequilibrium aspects of critical phenomena have
a wide range of important applications. Topical examples are
the active experimental searches and theoretical explorations of 
properties of the QCD critical point in the phase diagram of strongly 
interacting matter~\cite{QCDcritical}, or the description of the critical
dynamics of Bose-Einstein condensates in laboratory experiments
with ultracold quantum gases~\cite{BoseEinstein}. On extremely different 
energy scales their quantitative understanding requires 
a field theoretical description of the equilibrium and nonequilibrium
properties in the vicinity of critical points associated to second-order
phase transitions. The latter exhibit anomalously large fluctuations
and are characterized by universal quantities such as critical exponents.

There are few nonperturbative methods in thermal equilibrium that can 
describe the large fluctuations in the vicinity of a second-order phase 
transition~\cite{ZJ}. However, the number of methods
becomes even more limited once nonequilibrium dynamics and the
approach to thermal equilibrium is considered.
It is important to note that mean-field-type or leading-order
large-$N$ approximations are insufficient. They are
known to fail to describe thermalization and do not properly
distinguish the different universality classes for critical phenomena.
Efforts to go beyond leading order in a standard $1/N$ expansion of
the one-particle irreducible (1PI) effective action fail because
of spurious secular terms, which grow with time and invalidate
the approximation for nonequilibrium physics. 
In addition, the 1PI $1/N$ expansion shows a rather poor convergence 
in thermal equilibrium for
moderate values of $N$, and at low order important quantities such as the
critical anomalous dimension $\eta(N)$ are known to exhibit 
a spurious divergence as $N\to 0$~\cite{Ma:1973}.
  
It has recently been pointed out that a controlled description of
nonequilibrium dynamics and thermalization of quantum fields can be
based on a $1/N$ expansion of the two-particle irreducible (2PI)
effective action beyond leading
order~\cite{Berges:2001fi,Aarts:2002dj,Cooper:2002qd,Berges:2002cz,Berges:2002wr,Aarts:2004sd}.
For nonequilibrium dynamics the rapid convergence of this
expansion has also been observed in classical statistical field
theories, where comparisons with exact results are
possible~\cite{Aarts:2001yn}. This nonperturbative approach provides a
promising candidate for a uniquely suitable description of {\em both}
the nonequilibrium as well as equilibrium physics in the vicinity of
critical points, since it provides a controlled expansion even in the
presence of large fluctuations.  An important step towards such a
conclusion is, therefore, to show that the 2PI $1/N$ expansion indeed
reliably describes the thermal equilibrium properties at the critical
point of a second-order phase transition. In this work we calculate
universal properties, employing the 2PI $1/N$ expansion to
next-to-leading order (NLO).  We consider the $O(N)$-symmetric scalar
$N$-vector model in three-dimensions. For $N=4$ this model is expected
to belong to the universality class of the high temperature QCD phase
transition in the limit of two massless quark
flavors~\cite{Pisarski:ms}, and for $N=1$ to the QCD critical point at
high baryon density and temperature~\cite{Berges:1998rc}.  In the
context of Bose-Einstein condensation in quantum gases $N=2$
characterizes the relevant universality class~\cite{BoseEinstein}.
Here it is important to note that the equilibrium universality classes
agree in the relativistic and the nonrelativistic case, and we will
only consider the former.

To show the capabilities of the method we calculate
the properties of the theory directly at the critical
temperature of a second-order phase transition.
This is notoriously difficult within
perturbative approaches since the correlation lengths
are diverging. All the universal properties at this point
are encoded in a single critical exponent, which can 
be associated to the anomalous dimension $\eta$. In this case the 
knowledge of $\eta$ also fixes all universal information about the 
effective potential, which encodes the information 
about the critical equation of state.
We show that the 2PI $1/N$ expansion cures the spurious 
small-$N$ divergence of the standard (1PI) $1/N$ expansion,
and leads to improved estimates already for moderate values
of $N$. We finally compare the method with related 
approaches that have been employed in the literature.

\section{Model}

We consider a Euclidean field theory for a real, 
$N$--component scalar field 
$\varphi_a$ ($a=1,\ldots,N$) with classical action 
\beq
\label{eq:classical}
S[\varphi]=\int{\rm d}^{d}x
        \left(\frac{1}{2}\partial_\mu\varphi_a(x)\partial_\mu\varphi_a(x)  
        +\frac{m^2}{2}\varphi_a\!(x)\varphi_a\!(x) 
        +\frac{\lambda}{4!N}\left(\varphi_a\!(x)\varphi_a\!(x)\right)^2
 \right) \, ,
\eeq
where summation over repeated indices is implied.
All information about the equilibrium field theory 
can be obtained from the partition function, or more efficiently
from an $n$-particle irreducible effective action~\cite{Berges:2004pu}, 
which are related by Legendre transforms in the presence of sources. 
Here we consider the two-particle irreducible (2PI) 
representation of the effective action~\cite{Cornwall:1974vz}. 
The most general Euclidean 2PI effective action can be written as
\beq
\Gamma[\phi,G] = S[\phi] + \frac{1}{2} \Tr\ln G^{-1} 
          + \frac{1}{2} \Tr\, G_0^{-1}(\phi)\, G
          + \Gamma_2[\phi,G] +{\rm const} \, . 
\label{2PIaction}
\eeq
Diagrammatically the contribution $\Gamma_2[\phi,G]$ 
is given by all two-particle irreducible\footnote{A diagram is said 
to be two-particle irreducible if it does not become disconnected by 
opening two lines.} graphs with propagator lines set equal 
to $G$ \cite{Cornwall:1974vz}.
The classical inverse propagator $G_{0,ab}^{-1}(x,y;\phi)=
\delta^2 S[\phi]/\delta \phi_a(x) \delta \phi_b(y)$ reads
\bea
G^{-1}_{0,ab}(x,y;\phi) &=& \left( -\partial_\mu\partial_\mu + m^2 
+ \frac{\lambda}{6 N}\, \phi_c(x)\phi_c(x) \right) \delta_{ab}
\delta^{d}(x-y) \nonumber\\ 
&& + \frac{\lambda}{3 N}\, \phi_a(x) \phi_b(x) \delta^{d}(x-y) \, .
\label{classprop}
\eea
In the presence of an external source $J_a(x)$ 
coupling linearly to the fluctuating field, the equations of motion for
$\phi$ and $G$ are~\cite{Cornwall:1974vz} 
\bea
\frac{\delta \Gamma[\phi,G]}{\delta \phi_a(x)} = J_a(x) \quad , \quad 
\frac{\delta \Gamma[\phi,G]}{\delta G_{ab}(x,y)} &=& 0 \, .
\label{stationary}
\eea

\section{Two-particle irreducible $1/N$ expansion}
\label{2PINLOaction}

\begin{figure}[t]
\begin{center}
\includegraphics[width=0.9\textwidth]{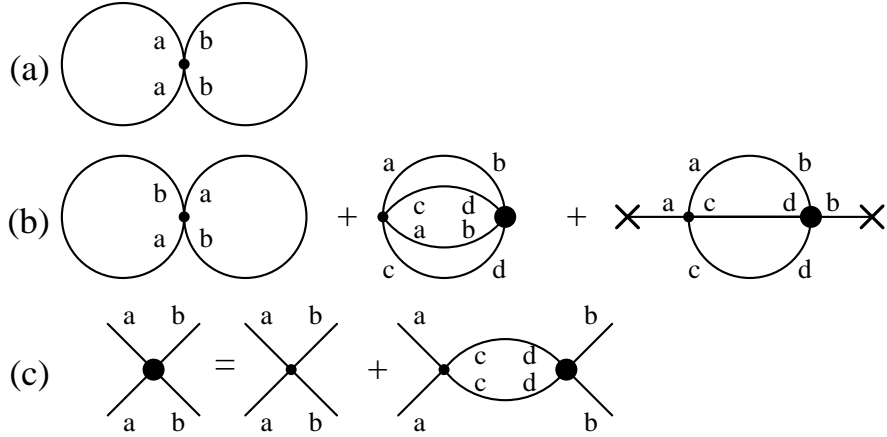}
\end{center}
\caption{Diagrammatic representation
  of the LO and NLO contributions to $\Gamma_2[\phi,G]$. The solid
  lines and small dots represent the full propagator and the bare
  vertex respectively and the indices run from $a=1,\ldots,N$.
  Diagram (a) is $\Gamma^{\rm LO}_2$ given by \eqn{LOcont}.  Diagrams
  (b) and (c) are $\Gamma^{\rm NLO}_2$, where we have expressed this
  contribution as a three-loop diagram with an effective four-vertex
  containing a ``chain'' of bubbles.  This form for $\Gamma^{\rm
  NLO}_2$ can been seen from \eqn{NLOcont} with the aid of
  \eqn{TraceLogB} and \eqn{ieqab}. }
\label{fig:action_diagram}
\end{figure}

In this work we consider a systematic expansion of the 2PI effective
action $\Gamma[\phi,G]$ in the number of field components or powers of
$1/N$ beyond leading order~\cite{Berges:2001fi,Aarts:2002dj}. We write \beq
\Gamma_2[\phi;G]= \Gamma_2^{\rm LO}[G] + \Gamma_2^{\rm NLO}[\phi;G] +
\Gamma_2^{\rm NNLO}[\phi;G] + \ldots \eeq where $\Gamma_2^{\rm LO}$
denotes the leading order (LO) contribution which scales proportional
to $N$, while $\Gamma_2^{\rm NLO}$ is the next-to-leading order (NLO)
contribution $\sim N^0$, and $\Gamma_2^{\rm NNLO}[G] \sim 1/N$ etc.
For the $O(N)$--model these contributions have been derived in
Ref.~\cite{Berges:2001fi,Aarts:2002dj}.\footnote{Note that in
  Ref.~\cite{Berges:2001fi,Aarts:2002dj} 
a Minkowskian space-time is considered,
  whereas here a Euclidean metric is used.}  The LO and NLO
contributions, which we will consider here, read 
(cf.~Fig.~\ref{fig:action_diagram})
\bea
\label{LOcont}
\Gamma_2^{\rm LO}[G] 
&=& \frac{\lambda}{4!N} \int_{x} G_{aa}(x,x)G_{bb}(x,x) \, ,
\\
\label{NLOcont} 
\Gamma_2^{\rm NLO}[\phi,G] &=&  \frac{1}{2} \Tr\,  
\mbox{ln} [\, {\bf B}(G)\, ]  \nonumber \\ 
&-& \frac{\lambda}{6N} \int_{xy} 
{\bf I}(x,y;G) \phi_a(x) G_{ab}(x,y) \phi_b (y). 
\eea
In the above equation we have defined
\beq
\label{Feq}
{\bf B}(x,y;G) = \delta^d(x-y)
 + \frac{\lambda}{6 N} G_{ab}(x,y)G_{ab}(x,y),
\eeq
and the logarithm in Eq.\ (\ref{NLOcont}) sums the infinite series 
\bea
\label{TraceLogB}
&&
\Tr\, \mbox{ln}[\, {\bf B}(G)\, ] 
=  \int_{x} 
 \left(\frac{\lambda}{6 N} G_{ab}(x,x)G_{ab}(x,x) \right)
\\
&& \quad -\frac{1}{2} \int_{xy}
\left(\frac{\lambda}{6 N} G_{ab}(x,y)G_{ab}(x,y) \right)
\left(\frac{\lambda}{6 N}\, G_{a'b'}(y,x)G_{a'b'}(y,x) \right)
+ \ldots \nonumber
\eea
The function ${\bf I}(x,y;G)$ is defined by 
\beq
{\bf I} (x,y;G) = \frac{\lambda}{6 N} G_{ab}(x,y) G_{ab}(x,y)
 - \frac{\lambda}{6 N} \int_{z} {\bf I}(x,z;G)
 G_{ab}(z,y) G_{ab}(z,y) \, ,
\label{ieqab}
\eeq
and resums an infinite number of ``chain'' graphs, which can be seen by 
iteratively expanding (\ref{ieqab}). 
The function ${\bf I} (x,y;G)$ and the inverse 
of ${\bf B}(x,y;G)$ are related by
\beq
{\bf B}^{-1}(x,y;G) = \delta^d(x-y) - {\bf I} (x,y;G) \, ,
\label{Binverse}
\eeq
which follows from convoluting Eq.\ (\ref{Feq}) with ${\bf B}^{-1}$ and
using Eq.\ (\ref{ieqab}). Note
that ${\bf B}$ and ${\bf I}$ do not depend
on $\phi$, and $\Gamma_2[\phi,G]$ is only quadratic in $\phi$ at NLO.

\section{Universal behavior at the critical point}

In the following we concentrate on $d=3$, relevant for
high temperature quantum field theories or classical 
statistical models with three spacial dimensions. For this case the
model is known to exhibit a second-order phase transition at a critical
mass parameter $m^2=m_c^2(N,\lambda)$ for all $N$.
In Fourier space we will use the notation
\mbox{$\int {\rm d}^3q/(2\pi)^3 \equiv \int_{\bq}\,$}.\footnote{If 
not stated otherwise most formulae are valid for general 
$d$ and the restriction to three dimensions will only be relevant
for some specific momentum integrals below.}
To show the capabilities of the method we calculate
the properties of the theory directly for the critical
value $m_c^2$. In this case the correlation lengths
are diverging, which spoils standard perturbative approaches. 
All the universal properties at this point
are encoded in a single critical exponent.\footnote{Note that 
deviations from the critical point are described by 
a further independent critical exponent.} In the absence
of external sources this exponent can be
associated to the anomalous dimension $\eta$. It 
characterizes the critical behavior of the propagator
or two-point function:    
\begin{equation}
  G(\bp) = \frac{1}{\bp^2}\left(\frac{\bp^2}{\Lambda^2}\right)^{\eta/2} \, ,
  \label{eq:prop_form}
\end{equation}
which is valid in the limit $\bp^2/\Lambda^2 \to 0^+$. Here $\Lambda$ 
corresponds to an (arbitrary) high momentum scale which
regularizes the theory. All universal properties are
independent of $\Lambda$ as is also shown below. 
 
We note that the knowledge of $\eta$ also fixes all 
universal information about the effective potential $U(\phi)$,
which encodes the complete information about the 
critical equation of state in the presence of an external 
(static) source $\sim J_a$ (cf.~(\ref{stationary}) and below). At the
critical point one has
\beq
\frac{\partial U(\phi)}{\partial \phi} \sim \phi^{\delta} \, ,
\label{eq:critU}
\eeq
valid for $\phi \to 0$ 
with a nonuniversal proportionality constant that depends on the specific 
details of the model. Here the universal critical exponent $\delta$
is not independent but related to $\eta$ by the scaling 
relation~\cite{ZJ}
\beq
\delta = \frac{5 - \eta}{1 + \eta}
\label{eq:scaling}
\eeq
for $d=3$. The presence of scaling relations such as
(\ref{eq:scaling}) is a consequence of the limited number
of relevant parameters, which can regulate the diverging
correlation lengths at the critical point. For the $O(N)$--model
there is only one independent relevant parameter at the
critical point for $m^2=m_c^2$, which can either be associated to
a nonvanishing momentum or source. As a consequence,
one may either extract the universal 
behavior from (\ref{eq:prop_form}) or from (\ref{eq:critU}).
We emphasize that the scaling relation 
(\ref{eq:scaling}) is a robust property of the theory in 
the absence of additional relevant parameters. 
In contrast, the error involved in an approximate estimate for a difficult
quantity such as the anomalous dimension is typically large.  
In the following we derive the relevant equations for both $\eta$
and $\delta$, employing the 2PI $1/N$ expansion to NLO. We will then
choose the simpler (former) case for an explicit solution of the
equations in order to obtain a quantitative estimate of the 
universal behavior.

Firstly, for a computation of the effective potential 
it is sufficient to consider a constant field expectation
value. By virtue of $O(N)$ rotations the most general field 
configuration in this case can be chosen as
\bea
\phi_a(x) &=& \sqrt{\frac{6N}{\lambda}}\, \phi\, \delta_{a1}  \, ,
\label{phiconf}
\\
G_{ab}(x,y) &=& {\rm diag}\left\{G_\pa(x-y),G_\pe(x-y),\ldots,G_\pe(x-y) 
\right\} \, ,
\label{Gconf}
\eea
where we have rescaled the field for later convenience. For $N>1$ the 
stationarity condition (\ref{stationary}) for the composite field $G$ with
(\ref{LOcont}) and (\ref{NLOcont}) 
translates into two coupled equations
for the longitudinal and transverse components. In Fourier space
one finds
\bea
G_\pa^{-1}(\bp) &=& \bp^2 + m^2 + 3 \phi^2 + \int_{\bq}
\Bigg\{
\frac{\lambda}{6N} \left[ 3 G_\pa(\bq) + (N-1) G_\pe(\bq) \right]
\nonumber\\
&-& 2 \phi^2\, {\bf I}(\bq) - \frac{\lambda}{3N} \left[
{\bf I}(\bq) + 2 \phi^2 G_\pa(\bq) \left(1 - {\bf I}(\bq)\right)^2 
\right] G_\pa(\bp-\bq)
\Bigg\} \, ,\nonumber\\[0.1cm]
G_\pe^{-1}(\bp) &=& \bp^2 + m^2 + \phi^2 + \int_{\bq}
\Bigg\{
\frac{\lambda}{6N} \left[ G_\pa(\bq) + (N+1) G_\pe(\bq) \right]
\nonumber\\
&-& \frac{\lambda}{3N} \left[
{\bf I}(\bq) + 2 \phi^2 G_\pa(\bq) \left(1 - {\bf I}(\bq)\right)^2 
\right] G_\pe(\bp-\bq)
\Bigg\} \, .
\label{gaphompe}
\eea 
Here the resummation function ${\bf I}$ is given by (\ref{ieqab})
for the configuration (\ref{Gconf}), in Fourier space. We will discuss 
the resummation function in more detail below.\footnote{The 
equation of motion for the field
expectation value is not required since one considers
$U(\phi)$ for all $\phi$. For completeness we note that 
the stationarity condition for the field (\ref{stationary}) leads 
for $\phi \not = 0$ to 
\beq
\phi^2 = -m^2 - \frac{\lambda}{6N}
\int \frac{{\rm d}^dq}{(2\pi)^d}\, 
\left[ 3 G_\pa(q) + (N-1) G_\pe(q) - 2\, {\bf I}(q) G_{\pa} (-q)\right] \, .
\eeq
}
The effective potential $U(\phi)$ is determined by the effective action
for a constant field:
\beq
U(\phi) = \frac{1}{V_3}\, \Gamma[\phi,G(\phi)]|_{\phi = {\rm const}} \,\, ,
\eeq 
where $V_3$ is the three-dimensional volume. 
Here $G(\phi) = \{G_\pa(\phi),
G_\pe(\phi) \}$ denotes the solutions of (\ref{gaphompe}).
We then obtain from (\ref{2PIaction}) with (\ref{LOcont})--(\ref{NLOcont})
the relevant equation for determining the equation of state:   
\bea
\frac{\partial U(\phi)}{\partial \phi} 
&=& \frac{6N\phi}{\lambda}\Big[ m^2 + \phi^2 + \frac{\lambda}{6N}
\int_{\bq} 
\Big\{ 3 G_\pa(\bq;\phi) + (N-1) G_\pe(\bq;\phi) 
\nonumber\\ 
&& - 2\, {\bf I}(\bq) G_{\pa}(-\bq;\phi)\Big\} \Big]\,  .
\label{eq:Ueos}
\eea   
We emphasize that the equations (\ref{gaphompe}) and (\ref{eq:Ueos}) 
are valid also away from the critical point of a second-order
phase transition. In particular, the effective potential can be
used to compute the different longitudinal and transverse 
susceptibilities in the phase with spontaneous symmetry 
breaking for $N>1$:
\beq
\chi_\pa^{-1} \sim \frac{\partial^2 U(\phi)}{\partial \phi\partial \phi}
\qquad , \qquad \chi_\pe^{-1} \sim \frac{1}{\phi}\, 
\frac{\partial U}{\partial \phi} \, .
\eeq
Note that Goldstone's theorem is fulfilled for the
2PI $1/N$ expansion, which has been pointed out previously 
in Ref.~\cite{Aarts:2002dj}. 
To compute the universal properties at the critical
point, i.e.~to obtain the exponent $\delta$ from (\ref{eq:critU}), 
one has to specify in (\ref{eq:Ueos}) the
critical value of the mass parameter $m^2 = m_c^2$. The latter can be obtained 
for $\phi = 0$ from either equation of (\ref{gaphompe}) by the condition 
$G^{-1}_{\pa}(\bp=0) \equiv G^{-1}_{\pe}(\bp=0) = 0$.

\begin{figure}[htb]
\begin{center}
\includegraphics[width=0.8\textwidth]{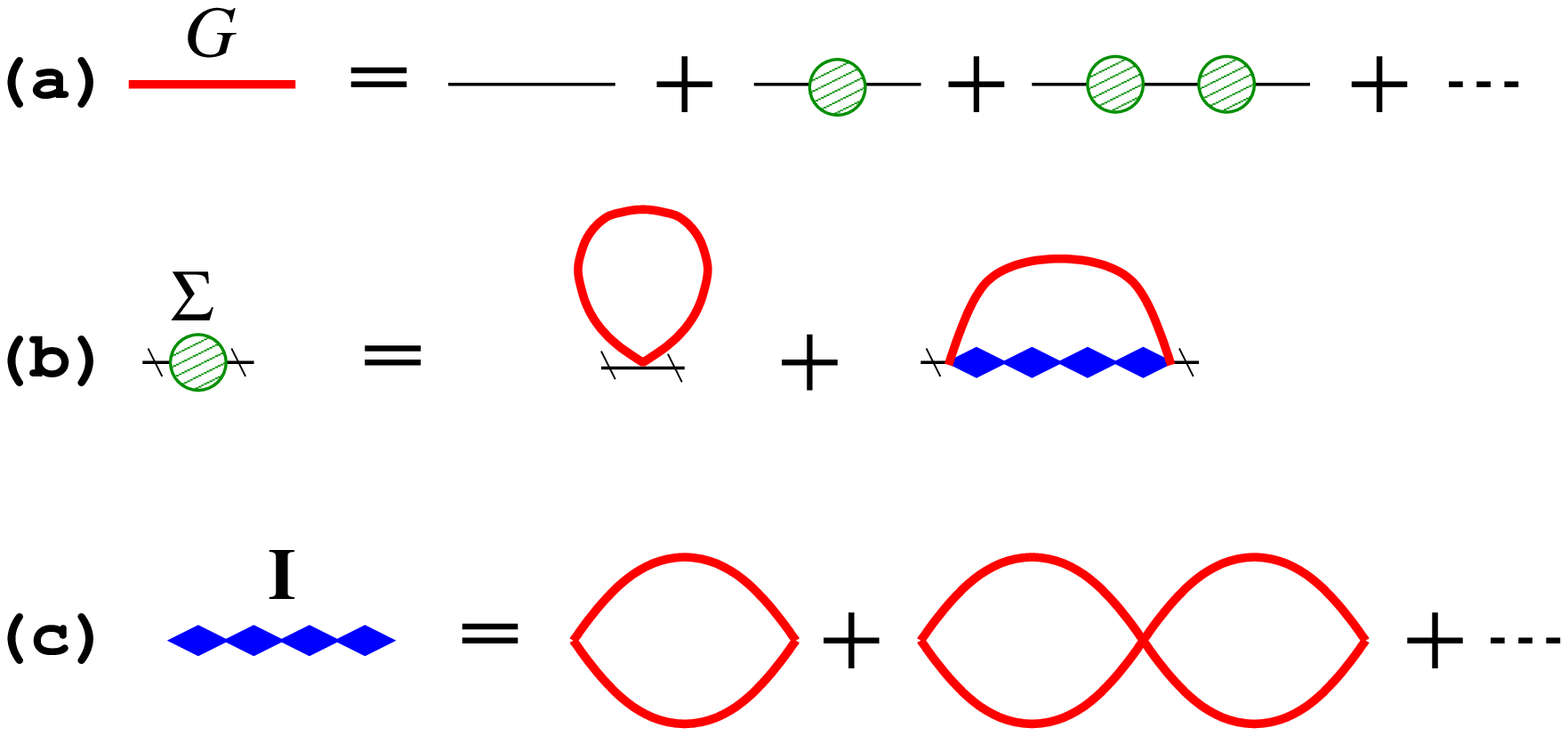}
\end{center}
\caption{Diagrammatic representation of self-consistent equations for the 
  full propagator $G(p)$ obtained from a $1/N$ expansion of
  the 2PI effective action to NLO for $\phi =0$. 
  Diagram (a) is \eqn{eq:gapfourier}, which
  expresses the full propagator in terms of the classical propagator
  $G_0$ (thin line) and the self-energy $\Si(p)$ (shaded blob).
  Diagram (b) is \eqn{eq:selfenergy}, which shows how the self-energy,
  the amputated one-particle-irreducible two-point function, can be
  expressed in terms of a one loop contribution containing the LO
  contribution, and a NLO contribution that involves the chain sum.
  Diagram (c), corresponding to \eqn{eq:explicitI}, shows how the
  chain sum can be expressed in terms of the full propagator.
}
\label{fig:diagram}
\end{figure}

An alternative calculation of the universal properties at the
critical point determines the anomalous dimension $\eta$. For our
current purposes it is sufficient to perform this calculation 
explicitly, which is a considerably simpler task.  
By virtue of $O(N)$ rotations, at the critical point  
the most general field configuration in the absence of sources
is given by
\beq
G_{ab}(\bp)=G(\bp)\, \delta_{ab} \, .
\label{eq:symconfig}
\eeq
Equations (\ref{gaphompe}) then reduce to the single expression
\beq
G^{-1}(\bp) = {\bp}^2 + m^2 + \Si(\bp)
\label{eq:gapfourier}
\eeq 
where the self-energy contains a momentum-independent ${\mathcal O}(N^0)$
part and both momentum-independent and momentum-dependent ${\mathcal O}(1/N)$ 
parts, 
\beq
\Si(\bp) = \lambda\, \frac{N+2}{6N} \int_{\bq} G(\bq) -
\frac{\lambda}{3N} \int_{\bq} G(\bp-\bq)\, {\bf I}(\bq) \,
\label{eq:selfenergy}
\eeq
and the chain sum is
\beq
{\bf I}(\bq) = 1- \left(1 + \frac{\lambda}{6}
\int_{\bk} G(\bq-\bk) G(\bk) \right)^{-1} \, . 
\label{eq:explicitI}
\eeq
The equivalence of the explicit form for the resummation function
(\ref{eq:explicitI}) with the implicit form of Eq.~(\ref{ieqab}) 
for the configuration (\ref{eq:symconfig}) 
can be observed by expansion in a geometric series. 
Eqs.~(\ref{eq:gapfourier})--(\ref{eq:explicitI}) provide a closed
set of self--consistent equations to determine the full propagator
$G(\bp)$. Fig.~\ref{fig:diagram} shows them in diagrammatic form.

According to (\ref{eq:prop_form}) at the critical point  
the inverse propagator at zero momentum vanishes, 
i.e.~$G^{-1}(\bp = 0) =0$. Using this and subtracting from
(\ref{eq:gapfourier}) the same expression for zero momentum we can write
\begin{equation}
G^{-1}(\bp)={\bp}^2 - \frac{\lambda}{3 N} \int_{\bq}
\left(G(\bp-\bq) - G(\bq)\right) {\bf I}(\bq) \, .
\end{equation}
This equation is valid at the critical point and can be 
conveniently employed to extract universal properties. 
With the help of (\ref{eq:explicitI}) this gives
\beq
G^{-1}(\bp)={\bp}^2 + \frac{\lambda}{3 N} \,
\int_{\bq}
\left(G(\bp-\bq) - G(\bq)\right)
\left(1 + \frac{\lambda}{6}
\int_{\bk}
 G(\bq-\bk) G(\bk) \right)^{-1} \, ,
\label{eq:gapcritical}
\eeq
where we used that the integral involving the constant part
of ${\bf I}(\bq)$ vanishes for $\bp^2 /\Lambda^2 \to 0$. The latter
limit characterizes the momenta for which the universal
exponent $\eta$ of the critical form 
for the propagator (\ref{eq:prop_form}) is defined. It remains
to be shown that (\ref{eq:prop_form})  
indeed solves the above equation
in this limit. To show this, we insert (\ref{eq:prop_form})
into (\ref{eq:gapcritical}) and obtain an equation for
the anomalous dimension $\eta$. The latter equation has
a unique solution for given $N$ as is demonstrated in the
following.

Using (\ref{eq:prop_form}) the one-loop subintegral in
Eq.~(\ref{eq:gapcritical}) can be 
performed for $-1 < \eta < 1/2$ to give\footnote{It 
can be conveniently obtained using the Feynman parametrization
for non-integer exponents $\alpha$, $\beta$:
\bea
\frac{1}{A^{\alpha}B^{\beta}} = \int_0^1 d x_1 d x_2\, \delta(x_1 + x_2 -1)
\frac{x_1^{\alpha-1} x_2^{\beta-1}}{\left( x_1 A + x_2 B \right)^{\alpha
+\beta}} \frac{\Gamma(\alpha+\beta)}{\Gamma(\alpha)\Gamma(\beta)} \, .
\nonumber
\eea
The momentum integration can then be performed for $\eta < 1/2$,
and the subsequent integration over the Feynman parameter for 
$\eta > -1$ with the result (\ref{eq:bubble_sum_integrand}).
Note that we have evaluated the integral without regulator.
One can show that subleading corrections obtained from 
keeping the momentum integration finite are irrelevant for the universal
low momentum behavior.  
}
\bea
  \label{eq:bubble_sum_integrand}
\frac{\lambda}{6} \int_{\bk} G(\bq-\bk) G(\bk) &=&
\frac{\lambda}{6 \Lambda}\, \int_{\bk/\Lambda} \,
\left(\frac{({\bf{q}}-{\bf{k}})^2}{\Lambda^2}\right)^{\eta/2-1}
\left(\frac{\bk^2}{\Lambda^2}\right)^{\eta/2-1}
\nonumber\\
&=& \frac{\lambda}{6\Lambda} 
\left(\frac{\bq^2}{\Lambda^2}\right)^{\eta - 1/2}
{\mathcal A}(\eta) \, ,
\eea
with
\beq
  {\mathcal A}(\eta) =\frac{1}{8 \pi^{3/2}}\,
  \frac{\Gamma(\frac{1}{2}-\eta)\left(\Gamma(\frac{1+\eta}{2})\right)^2}
  {\left(\Gamma(1-\frac{\eta}{2})\right)^2 \Gamma(1+\eta)} \, .  
  \label{eq:coeff_A}
\eeq
With this notation the equation (\ref{eq:gapcritical})
takes the form
\bea
\left(\frac{\bp^2}{\Lambda^2}\right)^{1-\eta/2} \!\!\!\! &=& 
 \frac{\bp^2}{\Lambda^2} + \frac{\lambda}{3 N \Lambda}\,  
\int_{\bq/\Lambda}
\left(\left(\frac{({\bf{p}}-{\bf{q}})^2}{\Lambda^2}\right)^{\eta/2-1} 
- \left(\frac{\bq^2}{\Lambda^2}\right)^{\eta/2-1}\right)
\nonumber\\
&\times &
\left(1+\frac{\lambda}{6\Lambda} 
\left(\frac{\bq^2}{\Lambda^2}\right)^{\eta - 1/2}
{\mathcal A}(\eta)\right)^{-1} \, .
  \label{eq:self_consist_1}
\eea
This provides a self-consistency equation for $\eta$ which may be solved
numerically for given $N$ and $\lambda/\Lambda$. We have done this
for a check which is discussed below. 

However, one can proceed further 
analytically to obtain $\eta$.     
For the low momentum range of critical phenomena, $\bp^2 /\Lambda^2 \to 0^+$,
the remaining momentum integral is dominated by small
$\bq^2 \sim \bp^2$ (cf.~above that \mbox{$\eta < 1/2$} for the allowed
range). In this limit we can, therefore, write
\beq
\left(\frac{\bp^2}{\Lambda^2}\right)^{1-\eta/2} \!\!\!= 
 \frac{\bp^2}{\Lambda^2} + \frac{2}{{\mathcal A}(\eta) N}
\int_{\bq/\Lambda}
\left(\!\left(\frac{({\bf{p}}-{\bf{q}})^2}{\Lambda^2}\right)^{\eta/2-1} 
\!\!\!- \left(\frac{\bq^2}{\Lambda^2}\right)^{\eta/2-1}\right)
\!\left(\frac{\bq^2}{\Lambda^2}\right)^{1/2 - \eta} \! .
  \label{eq:self_consist_2}
\eeq
One observes that the dependence on the coupling $\lambda$ dropped
out completely from the equation. This is a manifestation of
universality, which implies that the anomalous dimension for the
three-dimensional $O(N)$--model is only a function of the
number of field components $N$. 
After performing an elementary angle integration one finds
with $\eta \not = 0$ and using the notation 
${p} \equiv |\bp|/\Lambda$ and ${q} \equiv |\bq|/\Lambda$:  
\begin{equation}
{p}^{2-\eta} = 
{p}^2 + \frac{1}{2 \pi^2 {{\mathcal A}(\eta)} N}  
 \int_0^1 \!\! d{q} \: \left(
 \frac{{q}^{2-2\eta} }{\eta {p}}
\left(\! \left([{p}+{q}]^2\right)^{\eta/2} 
\!\!   - \left([{p}-{q}]^2\right)^{\eta/2} \right) 
-  2{q}^{1-\eta}
 \right)\!.
  \label{eq:self_consist_4}
\end{equation}
The remaining momentum integral can be performed with the 
help of hypergeometric functions. This is described in the
appendix. The result can be written as a sum of an anomalous term
$\sim {p}^{2-\eta}$ and a regular function $F_{\eta}(p^2)$ 
of momentum squared: 
\beq
{p}^{2-\eta} = p^2 + {\mathcal B}(\eta)\, {p}^{2-\eta} + F_{\eta}(p^2) \, ,
\label{eq:final}
\eeq
with
\bea
{\mathcal B}(\eta) &=&
\frac{4 \eta (1 - 2 \eta) \cos(\eta \pi)}{(3-\eta)(2-\eta) 
\sin^2(\eta \pi/2) N} \,+\, {\mathcal O}\left(\frac{1}{N^2}\right) \, .
\label{eq:coeff_constraints_1}
\eea
The regular function can be expanded in powers of the critical momentum
$p \to 0$ as
\beq
F_{\eta}(p^2) = - \frac{(1-\eta)(2-\eta)}{6 \pi^2 \eta {\mathcal{A}(\eta)} N}  
\, p^2 
+ {\mathcal O} \left(p^4, \frac{1}{N^2}\right) \, .
\label{eq:regF}
\eeq
Since the behavior of $F_{\eta}(p^2)$ for small momenta
is $\sim p^2$, this function along with the $p^2$-term in
(\ref{eq:final}) are subleading for a positive anomalous 
dimension. In this case, the $p \to 0$ behavior is dominated
by the anomalous term $\sim {p}^{2-\eta}$.
Comparing the left and the right hand side of Eq.~(\ref{eq:final}),
one observes that $\eta$ has to fulfill
\beq
{\mathcal B}(\eta) \stackrel{!}{=} 1 \, .
\label{eq:eta1}
\eeq
We observe from this constraint with (\ref{eq:coeff_constraints_1})
that the anomalous dimension indeed has to be positive for the allowed range 
$\eta < 1/2$, and (\ref{eq:eta1}) can be used to extract the
critical exponent $\eta$. In addition, we have numerically determined 
$\eta$ directly from Eq.~(\ref{eq:self_consist_1}) for selective
values of $N$, which agree.\footnote{We note that an approximate estimate 
for $\eta$ may also be obtained from enforcing that the
subleading terms $\sim p^2$ in (\ref{eq:final}) cancel, which leads
with (\ref{eq:regF}) to the condition
\bea
(1-\eta)(2-\eta) = 6 \pi^2 \eta {\mathcal{A}(\eta)} N \, . \nonumber
\eea
The solution of this equation and of (\ref{eq:eta1}) agrees to 
good accuracy for $N \gtrsim 1$, and have the same limit $\eta \to 0.5^-$ 
as $N \to 0^+$. However, only the leading contributions
for $p \to 0^+$ were calculated properly since they determine the critical 
behavior, and this is described by the solutions of (\ref{eq:eta1}).
} 

\begin{figure}
\begin{center}
\includegraphics[width=0.9\textwidth]{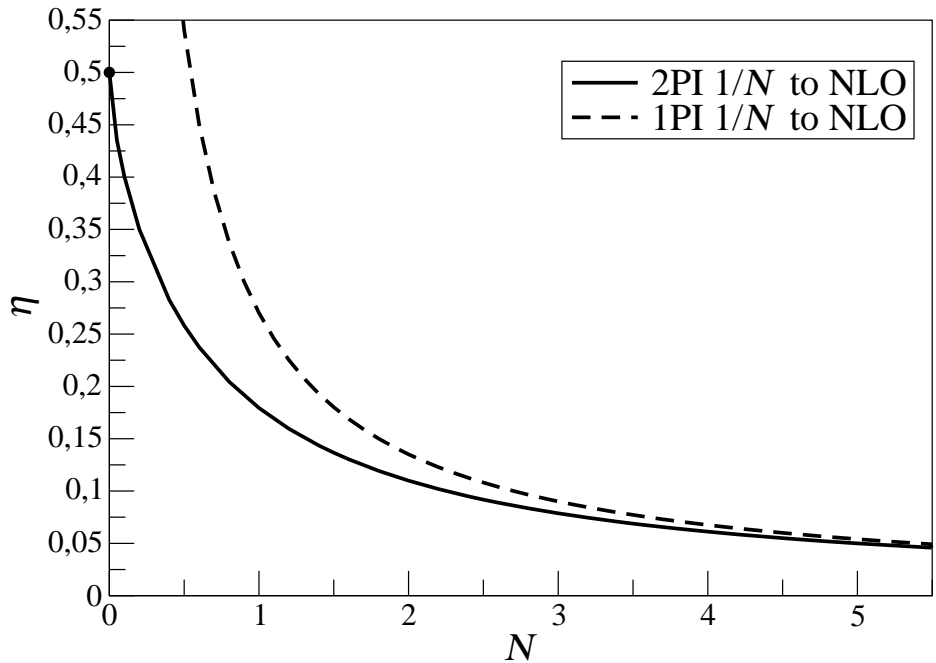}
\end{center}
\caption{The anomalous dimension $\eta$ for the three-dimensional
$O(N)$--model as a function of $N$. The figure compares the results
obtained from the 2PI $1/N$ expansion to next-to-leading order (NLO)
with the conventional 1PI $1/N$ expansion to NLO. One observes
that the expansion of the 2PI effective action cures the spurious 
$N \to 0$ divergence of the 1PI $1/N$ expansion.
This is reflected in the improved estimates of $\eta$ for small
$N$ if compared to results from alternative methods~\cite{ZJ}. 
In contrast, for large enough $N$ both expansions converge
to the same results and $\lim\limits_{N \to \infty} \eta(N) = 0$.}
\label{fig:eta2}
\end{figure}
In Fig.~\ref{fig:eta2} we have displayed the results for $\eta$
obtained from (\ref{eq:eta1}) as a function
of $N$ (solid line). For comparison, we also show the corresponding results
from the conventional (1PI) $1/N$ expansion to NLO 
(dashed line)~\cite{Ma:1973}. 
In contrast to the latter, one observes that the expansion of the
2PI effective action leads to a well-defined limit $\eta \to 0.5^-$ as
$N \to 0^+$.\footnote{The limit $N \to 0$ describes the universality class
for the critical swelling of long polymer chains~\cite{DeGennes}.} 
It therefore cures the spurious small-$N$ divergence of the 
standard (1PI) $1/N$ expansion, which is reflected in the improved
estimates of $\eta$ for moderate values of $N$. Despite this 
qualitative and quantitative improvement, the NLO approximation 
within the 2PI $1/N$ expansion 
still cannot compete with the more elaborate estimates from
alternative methods for small $N$. For instance, the 2PI result for $N=4$
is $\eta = 0.061$, which is still about $35$--$40$\% too high
compared to alternative estimates~\cite{ZJ}. 

In the following we compare the 2PI $1/N$ expansion with
previous approximations based on an ansatz for the
exact Schwinger-Dyson equation for the propagator.  
We note that the result from the 2PI $1/N$ expansion at NLO 
(\ref{eq:eta1}) agrees with the expression
obtained by Bray~\cite{Bray:1974} from his ``Self-Consistent Screening 
Approximation'' (SCSA). The latter approximation scheme corresponds
also to the so-called ``Bare Vertex Approximation'' (BVA), which
has been employed in the context of time evolution 
problems~\cite{Cooper:2002qd}. For BVA/SCSA one introduces an 
auxiliary field for a composite operator into the $O(N)$-model. 
In addition to the original propagator $G$, the
effective theory then contains a propagator for the composite
field and a two-point function mixing the original and
composite field. The exact Schwinger-Dyson equations for the 
two-point functions are then approximated by keeping the
interactions of the corresponding effective theory bare.
As has been pointed out in Ref.~\cite{Aarts:2002dj}, the BVA
is not consistent with the $1/N$ expansion of the 2PI effective 
action in the presence of a source term or for
$\phi \not = 0$, since BVA sums NLO and only part of the NNLO
contributions. For $\phi \equiv 0$, the 2PI $1/N$ expansion
to NLO and the BVA/SCSA are identical, which is the reason
for the agreement observed above. We emphasize that the 
2PI $1/N$ expansion can be systematically 
improved and going beyond this order requires
to include the NNLO corrections described in 
Ref.~\cite{Berges:2001fi,Aarts:2002dj}.

\section{Conclusions}

The 2PI $1/N$ expansion represents one of the
few nonperturbative methods that can calculate  critical behavior
directly in three dimensions. In particular, there are
no improvement procedures involved such as Borel transformation and
conformal mapping underlying results from expansions in $4-\epsilon$
dimensions~(cf.~the first Ref.~of~\cite{ZJ}). 
In view of the improved behavior of 
the 2PI $1/N$ expansion as compared to the standard (1PI) $1/N$ expansion,
the former seems to provide a promising candidate for  
quantitative estimates. Here it is important that
the $1/N$ expansion of the 2PI effective action can be
systematically improved. A very interesting further step would take 
into account the NNLO corrections as described in 
Ref.~\cite{Berges:2001fi,Aarts:2002dj}.
But already the NLO approximation provides a valuable
quantitative tool for studying critical phenomena. This concerns 
in particular real-time properties of quantum field theories.
It should be stressed that the 2PI $1/N$ expansion presents, 
so far, a uniquely suitable method that can deal with both 
nonequilibrium~\cite{Berges:2001fi,Aarts:2002dj,Cooper:2002qd,Berges:2002cz,Berges:2002wr,Aarts:2004sd} 
as well as equilibrium problems even 
in the presence of large fluctuations. Our results are very 
encouraging for an application of these methods to
dynamical properties of critical phenomena.

\vspace{3ex}
{\samepage 
\begin{center} Acknowledgements \end{center}
\nopagebreak 
The work of MGA was partially supported by the Department of Energy
under grant number DE-FG02-91ER40628. The work of JMC is supported by a
University Scholarship from Glasgow University, and additional funding
from a short-term DAAD scholarship, Glasgow University physics department
and Glasgow University Physical Sciences Graduate School.  JMC
acknowledges Washington University in St Louis and Institut f{\"u}r
Theoretische Physik, Universit{\"a}t Heidelberg for their hospitality
and thanks Greig Cowan for many helpful discussions.  }

\section{Appendix}
 
In this appendix we discuss the analytical 
evaluation of the momentum integral in Eq.~(\ref{eq:self_consist_4}).
Writing the equation as
\begin{equation}
  \label{eq:self_consist_5}
p^{2-\eta} = 
p^2 + \frac{1}{2 \pi^2 {{\mathcal A}(\eta)} N} 
 \left(R_1 + R_2 +R_3 \right) \, 
\end{equation}
we decompose the integral into three terms. The first integral is
\begin{equation}
  \label{eq:int_1a}
 R_1= \frac{p^{\eta-1}}{\eta} \int_0^1 dq \: q^{2-2\eta}
\left(1+\frac{q}{p}\right)^{\eta} \, ,
\end{equation}
which in terms of the hypergeometric function 
\begin{equation}
  \label{eq:HyperGeo}
  \phantom{1}_2F_1(a,b,c,z) = \frac{\Gamma(c)}{\Gamma(b)\Gamma(c-b)}
\int_0^1 dx \: \frac{x^{b-1}(1-x)^{c-b-1}}{
\left(1-x z\right)^{a}}
\end{equation}
becomes
\bea
  \label{eq:int_1b}
 R_1 &=& 
\frac{p^{\eta-1}}{\eta} \frac{1}{3-2\eta}
 \phantom{1}_2F_1\left(3-2\eta,-\eta,4-2\eta,-\frac{1}{p}\right) \, ,
\eea
where we have simplified products of $\Gamma$-functions and relied upon the
relation 
\begin{equation}
  \label{eq:hypergeo_arrange}
  \phantom{1}_2F_1(a,b,c,z) = \phantom{1}_2F_1(b,a,c,z).  
\end{equation}
The hypergeometric function can be transformed further 
using~\cite{Gradshteyn:1994} 
\begin{eqnarray}
  \label{eq:HypGeoTrans1} \hspace*{-1cm}
\phantom{1}_2F_1\left(a,b,c,z\right)\!\!\! &=& \!\!\!
  \frac{\Gamma(c)\,\Gamma(b-a)}{\Gamma(b)\,\Gamma(c-a)}\, (-z)^{-a} 
   \phantom{1}_2F_1\left(a,1+a-c,1+a-b,\frac{1}{z}\right) \\ \nonumber
   &+& \!\!\!
   \frac{\Gamma(c)\,\Gamma(a-b)}{\Gamma(a)\,\Gamma(c-b)}\, (-z)^{-b} 
   \phantom{1}_2F_1\left(b,1+b-c,1+b-a,\frac{1}{z}\right) \, .
\end{eqnarray}
This yields hypergeometric functions suitable for a small
$p$ expansion,
i.e.
\bea
\lefteqn{ \!\!\!\!\!\!\!\!
\phantom{1}_2F_1\left(3-2\eta,-\eta,4-2\eta,
-\frac{1}{p}\right)  
\, = \,  p^{3-2\eta}\,
      \left( \frac{\Gamma(4-2\eta)\, \Gamma(-3 + \eta)}
         {\Gamma(-\eta)} \right) } \qquad\qquad \nonumber \\  
&& +\, p^{-\eta}\,\Bigg( \frac{3-2\eta}{3-\eta} 
+
        \frac{\eta\, (3-2\eta) }{\left( 2 - \eta \right)}\, p  
-   
    \frac{\eta\, (3-2\eta) }{2\,}\, p^2 \nonumber\\  
&& - \,     
    \frac{(3-2\eta) \,\left( 1 -\eta \right) \,\left( 2 -\eta \right)}
    {6 }\, p^3 
+  {\mathcal O}\left(p^4\right)\! \Bigg) .
  \label{eq:HypGeoExpand_part1} 
\eea
The second integral is 
\begin{equation}
  \label{eq:int_2a}
R_2= -\frac{p^{\eta-1}}{\eta} \int_0^1 dq \: q^{2-2\eta}
\left(\left[1-\frac{q}{p}\right]^2\right)^{\eta/2} \, ,
\end{equation}
which may be evaluated with
the aid of the following relation where $z>1$:
\begin{eqnarray}
  \label{eq:HypGeoRelation}
 \lefteqn{ \int_0^1dx\:x^a(1-xz)^{2b} = z^{-1-a}
\Bigg(\frac{\Gamma(-1-a-2b)\Gamma(1+2b)}{\Gamma(-a)} } \\
&\!\!\! +&\!\!\! \frac{\Gamma(1+a)\Gamma(1+2b)}{\Gamma(2+a+2b)} \Bigg) 
+ \frac{z^{2b}}{(1+a+2b)} \phantom{1}_2F_1(-1-a-2b,-2b,-a-2b,1/z) \, .
\nonumber
\end{eqnarray}
So we obtain
\begin{eqnarray}
R_2 &=&  -\frac{p^{2-\eta}}{\eta} 
  \left(\frac{\Gamma(-3+\eta)\Gamma(1+\eta)}{\Gamma(-2+2\eta)} 
  + \frac{\Gamma(3-2\eta)\Gamma(1+\eta)}{\Gamma(4-\eta)} \right) \\ \nonumber
& -& \frac{1}{\eta p} \frac{1}{(3-\eta)} 
\phantom{1}_2F_1(-3+\eta,-\eta,-2+\eta,p)  \, .
\end{eqnarray}
Expanding the hypergeometric function in small
$p$ one finds 
\begin{eqnarray}
\lefteqn{
  \phantom{1}_2F_1\left(-3+\eta,-\eta,-2+\eta,p \right)
  = 1 - \frac{\eta\,(3-\eta)}{2-\eta}\, p } \nonumber\\
  &-& 
  \frac{\eta\,(3-\eta)}{2} \, p^2 
+ \frac{(3-\eta) \,\left( 1 -\eta \right) \,\left( 2 -\eta \right)}{6}
  \, p^3 
+ {\mathcal O}\left( p^4 \right) \, .
\end{eqnarray}
The third integral is straightforward and gives  
\begin{equation}
  \label{eq:int_3}
R_3 = - \int_0^1 dq \: 2 q^{1-\eta} =  
 \frac{-2}{2-\eta} \, .
\end{equation}
Collecting the results one observes that terms 
containing odd powers of $p$
cancel
and up to ${\mathcal O}\left( p^4 \right)$
one finds (\ref{eq:final}) with 
(\ref{eq:coeff_constraints_1}) and (\ref{eq:regF}).

\pagebreak

\end{document}